\documentclass[floatfix, aps,prb,reprint,twocolumn,superscriptaddress,amsmath,amssymb]{revtex4-2}

\usepackage{graphicx}
\usepackage{dcolumn}
\usepackage{bm}
\usepackage{hyperref}
\hypersetup{
    colorlinks=true,
    linkcolor=black,   
    citecolor=blue,    
    urlcolor=cyan      
}

\newcommand{\ket}[1]{\lvert #1 \rangle}

\begin{document}

\preprint{APS/123-QED}

\title{Order-Dependent Modification of High-Harmonic Generation by Quantum Dissipation and Lamb Shift}

    \author{Xiangyu Zhang}
    \affiliation{Institute of Ultrafast Optical Physics, MIIT Key Laboratory of Semiconductor Microstructure and Quantum Sensing, Department of Applied Physics, Nanjing University of Science and Technology, Nanjing 210094, China}
    \affiliation{State Key Laboratory of Precision Spectroscopy, East China Normal University, Shanghai, 200241, China.}
    \author{Tong Wu}
    \affiliation{State Key Laboratory of Precision Spectroscopy, East China Normal University, Shanghai, 200241, China.}
    \author{Xiang Li}
    \affiliation{State Key Laboratory of Precision Spectroscopy, East China Normal University, Shanghai, 200241, China.}
    \author{Jianshi Lu}
    \affiliation{State Key Laboratory of Precision Spectroscopy, East China Normal University, Shanghai, 200241, China.}
    \author{Zhe Yang}
   \affiliation{State Key Laboratory of Precision Spectroscopy, East China Normal University, Shanghai, 200241, China.}
   \author{Chao Yu} \email{chaoyu@njust.edu.cn}
   \affiliation{Institute of Ultrafast Optical Physics, MIIT Key Laboratory of Semiconductor Microstructure and Quantum Sensing, Department of Applied Physics, Nanjing University of Science and Technology, Nanjing 210094, China}
    \author{Shicheng Jiang} \email{scjiang@lps.ecnu.edu.cn}
   \affiliation{State Key Laboratory of Precision Spectroscopy, East China Normal University, Shanghai, 200241, China.} 
   \author{Ruifeng Lu} \email{rflu@njust.edu.cn}
   \affiliation{Institute of Ultrafast Optical Physics, MIIT Key Laboratory of Semiconductor Microstructure and Quantum Sensing, Department of Applied Physics, Nanjing University of Science and Technology, Nanjing 210094, China}

\date{September 21, 2026}

\begin{abstract}
Many-body environments are conventionally incorporated into quantum dynamics as heat baths, which induce a Lamb shift and quantum dissipation. In the traditional picture, such environmental coupling is expected to induce suppression and broadening of spectral signals. In this letter, we investigate high-order harmonic generation (HHG) in a two-level system coupled to a heat bath via dipole-dipole interactions using the Lindblad master equation. It is found that the environmental effect does not simply suppress the harmonic efficiency. Instead, when the cutoff frequency of the coupling spectral density exceeds the energy-level spacing, the environment can actually enhance the harmonic yield. Further analysis reveals that this enhancement originates from intense level fluctuations induced by the Lamb shift. Meanwhile, the environmental influence exhibits a distinct dependence on harmonic order, manifesting clearly different behaviors for lower and higher harmonics. Our results challenge the common relaxation-time picture of uniform damping and establish a microscopic, order-dependent mechanism for environmental control of HHG, with direct implications for solid-state attosecond spectroscopy and quantum material engineering.
\end{abstract}

\maketitle



High-order harmonic generation (HHG) is a cornerstone of strong-field physics and attosecond science, observed as coherent radiation at integer multiples of the incident frequency when intense ultrafast lasers interact with matter~\cite{McPherson1987, Ferray1988}. The underlying physics is elegantly captured by the single-active-electron three-step recollision model~\cite{Corkum1993, Lewenstein1994}, which has enjoyed remarkable success in describing HHG in atomic and molecular gases. This single-particle picture has been extended to a wide range of condensed-phase systems, including bulk crystals~\cite{Ghimire2011}, organic molecular crystals~\cite{Wiechmann2025}, two-dimensional materials~\cite{Spintextures2025,Monolayer2025}, liquids~\cite{Luu2018,Mondal2025}, and laser-induced plasmas~\cite{Kim2022,Ganeev2023}. These developments have positioned HHG as a versatile platform for probing ultrafast dynamics and quantum phenomena, enabling applications such as isolated attosecond pulse generation~\cite{Vismarra2023, Guo2026}, electronic structure retrieval~\cite{Parks2025, Banks2017}, frequency comb generation~\cite{Rutledge2021, Di2023}, non-classical light synthesis~\cite{Gorlach2020, Stammer2024,jiangCPL}, and terahertz technology~\cite{TopoMeta2025}.

In solids, however, the single-particle picture faces a fundamental limitation: many-body interactions such as electron-phonon coupling and the dipole-dipole interaction between excitons cannot be ignored. Electron–hole coherence driven by the laser field is inherently dissipative due to many-body interactions~\cite{Brown2024, Bae2026, Hu2024, Herling2026}. To account for these effects, most theoretical studies have adopted a phenomenological approach, introducing parameterized dephasing on the order of a few femtoseconds or even shorter~\cite{Liu2025, Frontiers2025, Discreteness2023}. In this conventional picture, dephasing and relaxation are expected to simply suppress harmonic yields and broaden spectral lines—a purely destructive role for the environment.

Is this the full story? Recent work has begun to challenge this view by employing open quantum system frameworks that go beyond simple parameterized damping ~\cite{UltrafastDephasing2023, Boroumand2025}. The Lindblad master equation has emerged as a powerful tool for describing dissipative electron dynamics in HHG~\cite{Bae2026, Reentrance2026, Decoherence2021}. Non-Markovian effects have been explored to avoid overestimating charge-carrier populations~\cite{Incoherent2023, Kruchinin2019}. Temperature-induced dephasing has been shown to affect harmonic generation in a nontrivial manner~\cite{Du2022, Trajectory2024}. Spin–phonon coupling in laser-driven systems has further highlighted the importance of a microscopic description of the environment for HHG control~\cite{SpinHHG2024}. These studies point toward a more complete picture: the environment may play a role far more complex than mere suppression.

In this letter, we develop a microscopic model of a driven two-level system coupled to a heat bath via dipole-dipole interactions, using the Lindblad master equation~\cite{Lindblad1976, Gorini1976}. Our key innovation is to decompose the environmental effect into two distinct components—a Lamb shift that renormalizes the energy levels, and quantum dissipation that accounts for dephasing and population transfer~\cite{Breuer2002, Leggett1987}—and to study their separate roles in HHG. We find that the environment is not merely destructive: when the cutoff frequency of the coupling spectral density exceeds the level spacing, the Lamb shift can actually enhance harmonic emission. This enhancement is order-dependent—low harmonics respond monotonically, while high harmonics exhibit nonmonotonic behavior as dissipation competes with the Lamb-shift-induced gain. We trace the enhancement mechanism to a frequency-doubling modulation of the energy levels that preferentially opens sideband channels for higher harmonics. Our results challenge the common relaxation- and dephasing-time picture of uniform damping in the strong-field regime.

\begin{figure}
  \centering
  \includegraphics[width=\columnwidth]{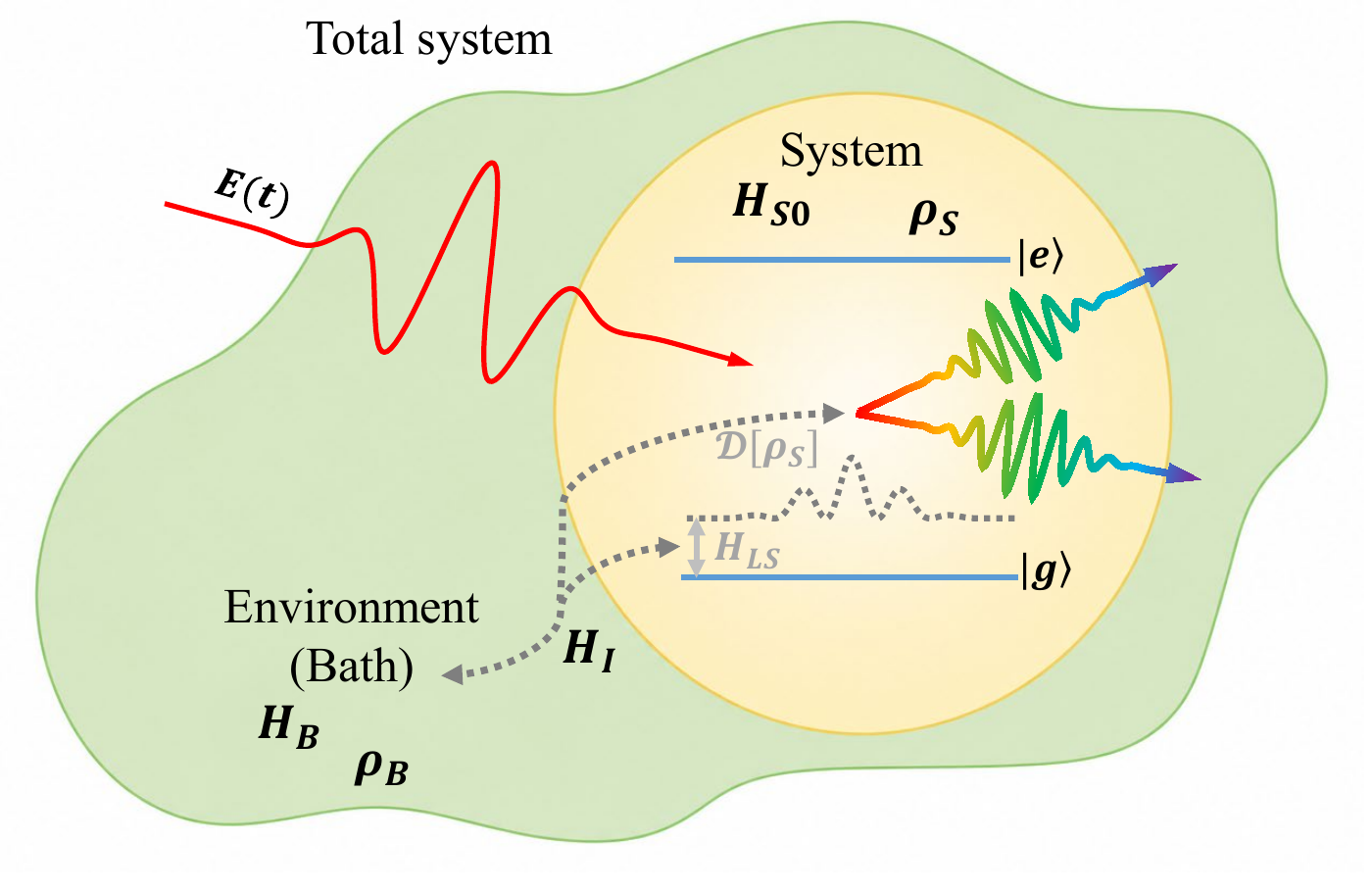}
  \caption{(Color online) Schematic illustration of the total system. The two-level system with ground state $\ket{g}$ and excited state $\ket{e}$ is driven by a strong laser pulse and coupled to a heat bath via dipole-dipole interactions. The total Hamiltonian is decomposed into the system part $H_{S0}$, the bath part $H_B$, and the interaction $H_I$. After tracing out the bath degrees of freedom, the reduced density matrix $\rho_S$ of the system evolves according to the Lindblad equation, where the environmental effect is separated into a Lamb shift Hamiltonian $H_{\mathrm{LS}}$ (renormalizing the energy levels) and a dissipator $\mathcal{D}[\rho_S]$ (accounting for dephasing and population transfer). The bath is assumed to remain in a thermal state $\rho_B$ at temperature $T$.}
  \label{fig:schematic}
\end{figure}

The open quantum system is illustrated schematically in Fig.~\ref{fig:schematic}. A two-level electronic system is driven by a strong laser field and coupled to a heat bath modeled by harmonic oscillators~\cite{Caldeira1983} via dipole-dipole interactions. The total Hamiltonian is decomposed into the system part \(H_S\), the bath part \(H_B\), and the interaction \(H_I\) between the system and the environment (atomic units are adopted unless explicitly stated):
\begin{equation}
H_T = H_S(t) + H_B + \alpha H_I,
\label{eq:total_H}
\end{equation}
where \(\alpha\) measures the overall interaction strength. The system Hamiltonian is \(H_S(t) = H_{S0} + Q_S \cdot E(t)\), where \(H_{S0}\) is the field-free Hamiltonian of the two-level system, \(E(t)\) is the external laser field, and \(Q_S\) is the dipole operator. The bath Hamiltonian is \(H_B = \sum_k \omega_k b_k^\dagger b_k\), where \(b_k^\dagger\) and \(b_k\) are the creation and annihilation operators of the \(k\)-th bath mode and \(\omega_k\) is its frequency. The interaction takes the dipole-dipole form \(H_I = Q_S \sum_k g_k q_k\), where \(q_k = (b_k + b_k^\dagger)/\sqrt{2}\) are dimensionless bath position operators and \(g_k\) are the coupling constants.

We assume that initially the system and the bath are uncorrelated and the bath is in thermal equilibrium~\cite{Breuer2002}, with \(\rho_T(0) = \rho_S(0) \otimes \rho_B\), where \(\rho_T(0)\), \(\rho_S(0)\), and \(\rho_B\) denote the initial total density operator, the initial reduced density operator of the system, and the thermal equilibrium density operator of the bath, respectively. The system-bath interaction is weak (\(\alpha \ll 1\)), and the system-bath correlation time is much shorter than the characteristic evolution time of the system. Under these conditions, by applying the Born-Markov approximation~\cite{Breuer2002,Gardiner2000}, tracing out the bath degrees of freedom yields a Lindblad master equation for the reduced density matrix \(\rho_S(t) = \operatorname{Tr}_B[\rho_T(t)]\). The result is
\begin{equation}
\frac{\partial \rho_S}{\partial t} = -i\bigl[H_S(t) + H_{LS}(t),\; \rho_S\bigr] + \mathcal{D}[\rho_S],
\label{eq:lindblad}
\end{equation}
where the Lamb shift Hamiltonian is
\begin{equation}
H_{LS}(t) = \sum_{\omega(t)} \pi(\omega(t),t) \, Q_S^\dagger(\omega(t)) Q_S(\omega(t)),
\label{eq:Lamb_shift}
\end{equation}
and the dissipator reads
\begin{align}
\mathcal{D}[\rho_S] = \sum_{\omega(t)} \gamma(\omega(t),t) \Bigl[ & Q_S(\omega(t)) \rho_S Q_S^\dagger(\omega(t)) \notag \\
& - \frac{1}{2}\bigl\{ Q_S^\dagger(\omega(t)) Q_S(\omega(t)), \rho_S \bigr\} \Bigr],
\label{eq:dissipator}
\end{align}
with \(Q_S(\omega(t))\) being the expansion of the dipole operator in the eigenbasis of the time-dependent Hamiltonian, where \(\omega(t)\) is the corresponding eigenvalue. The coefficients \(\gamma(\omega(t),t)\) and \(\pi(\omega(t),t)\) are the dissipation rate and the Lamb shift coefficient, determined by the bath correlation function. The bath is characterized by its spectral density \(J(\omega_1) = \pi/2 \sum_k g_k^2 \delta(\omega_1 - \omega_k)\), where \(\omega_k\) is the frequency of the \(k\)-th bath mode. In this letter we adopt the Ohmic spectral density with an exponential cutoff, \(J(\omega_1) = \eta \, \omega_1 \, e^{-\omega_1/\omega_c}\), where \(\eta\)(\(\propto\alpha^2\)) is the effective overall coupling strength and \(\omega_c\) is the cutoff frequency of the interaction spectral density. A detailed derivation of the Lindblad master equation and the explicit forms of \(\gamma(\omega(t),t)\) and \(\pi(\omega(t),t)\) in terms of \(J(\omega_1)\) are given in the Supplemental Material~\cite{Supplemental}.

Equations \eqref{eq:lindblad}--\eqref{eq:dissipator} constitute the main theoretical framework of this letter. They show how the environment influences the system through a renormalization of the energy levels (Lamb shift) and through quantum dissipation that includes both dephasing and population transfer. The parameters \(\omega_c\) and \(\eta\) appear explicitly in the rates \(\gamma(\omega(t),t)\) and the Lamb shift coefficients \(\pi(\omega(t),t)\), giving rise to the parameter dependences studied in the following sections. No further approximations are made in the numerical solution of Eq.~\eqref{eq:lindblad}; the time-dependent Hamiltonian and the time-dependent coefficients are treated exactly.

We consider a two-level system with energy gap $\Delta E = 9$ eV which is close to the band gap of SiO$_2$, driven by a laser pulse of wavelength $\lambda = 1600$ nm with a Gaussian envelope of $ 50$ fs full width at half maximum. The heat bath is maintained at $T = 300$ K. The harmonic spectra are computed for orders $n = 5, 7, 9, 11, 13, 15$ by solving the Lindblad master equation derived above. To quantify the environmental effects, we define the ratio $R_n = I_n/I_n^{(0)}$, where $I_n$ is the harmonic intensity in the presence of the bath and $I_n^{(0)}$ is that of the isolated system. This ratio is mapped as a function of the cutoff frequency $\omega_c$ of the interaction spectral density (scanned from $0.1\Delta E$ to $10\Delta E$) and the system-bath coupling strength $\eta$ (scanned from $10^{-4}$ to $10^{-2}$).

\begin{figure}
  \centering
  \includegraphics[width=\columnwidth]{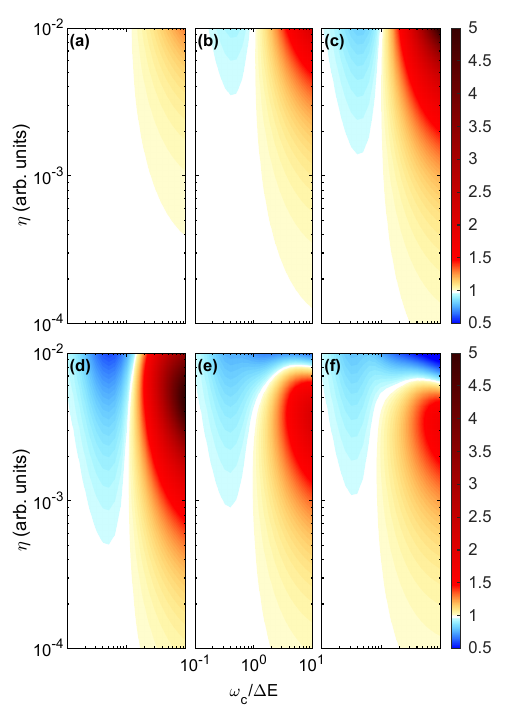}
  \caption{(Color online) Ratio of the harmonic intensity in the presence of the bath to that without the bath, $R_n = I_n/I_n^{(0)}$, as a function of the cutoff frequency $\omega_c$ of the interaction spectral density (horizontal axis, in units of $\Delta E$) and the system-bath coupling strength $\eta$ (vertical axis). Panels (a)--(f) correspond to harmonic orders 5, 7, 9, 11, 13, and 15, respectively. The energy gap $\Delta E$ is slightly larger than the 11th harmonic order. The color scale indicates enhancement (red, $R_n > 1$) or suppression (blue, $R_n < 1$).}
  \label{fig:full}
\end{figure}

The computed harmonic intensity ratios $R_n$ are shown in Fig.~\ref{fig:full} (panels a--f for orders 5, 7, 9, 11, 13, and 15, respectively). A clear threshold behavior is observed at $\omega_c \approx \Delta E$ for all harmonic orders. When $\omega_c < \Delta E$, all harmonics are suppressed ($R_n < 1$, blue regions). When $\omega_c > \Delta E$, the response splits into two distinct classes. For low-order harmonics (5, 7, 9; Figs.~\ref{fig:full}(a)--\ref{fig:full}(c)), the intensity becomes enhanced ($R_n > 1$, red regions) for $\omega_c > \Delta E$. The enhancement grows monotonically with both increasing $\omega_c$ and increasing $\eta$. For high-order harmonics (11, 13, 15; Figs.~\ref{fig:full}(d)--\ref{fig:full}(f)), the behavior is qualitatively different. While suppression for $\omega_c < \Delta E$ is again observed, the region $\omega_c > \Delta E$ exhibits a strong nonmonotonic dependence on $\eta$: for small $\eta$, the intensity is enhanced; as $\eta$ increases, the enhancement reaches a maximum and then declines, and the harmonic intensity gradually turns from enhancement to suppression. Within the range of $\eta$ explored here, this nonmonotonic behavior is not observed for low-order harmonics.

\begin{figure}
  \centering
  \includegraphics[width=\columnwidth]{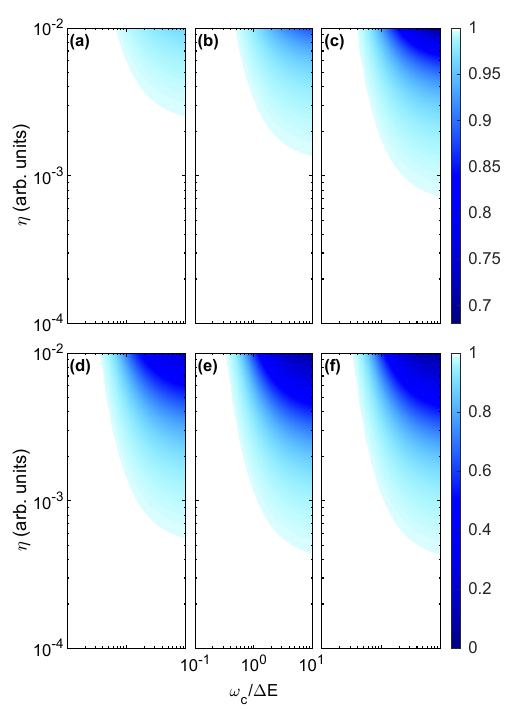}
  \caption{(Color online) Same as Fig.~\ref{fig:full}, but with only the quantum dissipation contribution retained (the Lamb shift Hamiltonian is omitted). This corresponds to the effects of dephasing and population transfer induced by the environment. Panels (a)--(f) correspond to harmonic orders 5, 7, 9, 11, 13, and 15, respectively.}
  \label{fig:diss}
\end{figure}

\begin{figure}
  \centering
  \includegraphics[width=\columnwidth]{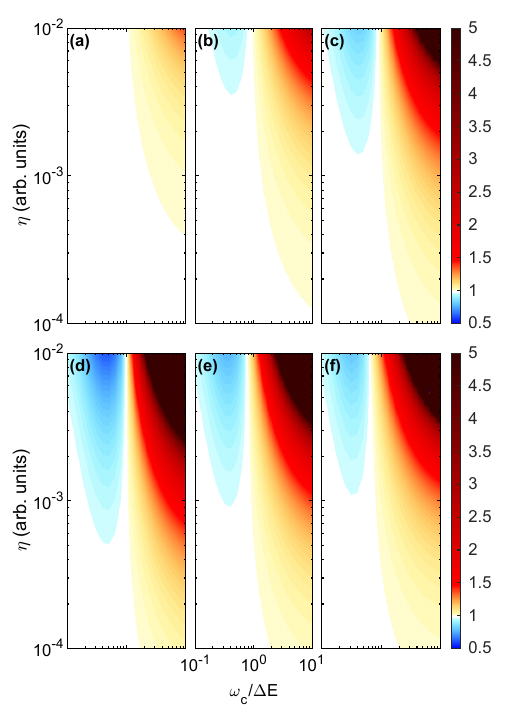}
  \caption{(Color online) Same as Fig.~\ref{fig:full}, but with only the Lamb shift contribution retained (the dissipative part of the Lindblad equation is set to zero). This corresponds to the effect of energy-level renormalization by the environment. Panels (a)--(f) correspond to harmonic orders 5, 7, 9, 11, 13, and 15, respectively.}
  \label{fig:lamb}
\end{figure}

As shown in the Lindblad equation [Eq.~\eqref{eq:lindblad}], the environmental effect on the system naturally separates into two distinct contributions: the Lamb shift (energy renormalization) and quantum dissipation (dephasing and population transfer). To understand how these two contributions affect the harmonic emission and give rise to the phenomena observed in Fig.~\ref{fig:full}, we perform two additional sets of calculations: in one, only the dissipator is kept; in the other, only the Lamb shift is retained. The results are shown in Figs.~\ref{fig:diss} and \ref{fig:lamb}, respectively.

\textbf{Dissipation only (Fig.~\ref{fig:diss}).} When only the dissipator is kept, the harmonic intensity is reduced for all $\omega_c$ and $\eta$, with the reduction stronger for higher harmonics and larger $\eta$. No enhancement is observed in any parameter region. This indicates that quantum dissipation alone cannot produce any of the enhancement seen in Fig.~\ref{fig:full}; it only suppresses harmonic emission.

\textbf{Lamb shift only (Fig.~\ref{fig:lamb}).} When only the Lamb shift is retained, the threshold at $\omega_c \approx \Delta E$ is fully reproduced: the intensity is suppressed for $\omega_c < \Delta E$ and enhanced for $\omega_c > \Delta E$. Across the entire parameter range, the effect of the Lamb shift grows monotonically with increasing $\eta$. This confirms that the $\omega_c$-dependent switching behavior in Fig.~\ref{fig:full} originates from the Lamb shift.

The observation that the Lamb-shift-induced enhancement is significantly stronger for higher-order harmonics calls for a deeper understanding of its frequency characteristics. According to Eq.~\eqref{eq:Lamb_shift}, the time dependence of the Lamb shift originates entirely from the instantaneous eigenfrequency $\omega(t) = \pm\sqrt{(\Delta E)^2 + 4|d|^2|E(t)|^2}$ of the time-dependent system Hamiltonian, so its fluctuation spectrum contains only even multiples of the laser frequency. This property is directly inherited by the Lamb shift.

\begin{figure}
    \centering

    \newlength{\labelwd}\setlength{\labelwd}{0.4cm}
    \newlength{\imgwd}\setlength{\imgwd}{\dimexpr\columnwidth-2\labelwd\relax}

    \newlength{\imgh}\setlength{\imgh}{\dimexpr\imgwd*3/4\relax}

    \begin{minipage}{\columnwidth}
        \centering
        \begin{minipage}[c]{\labelwd}
            \rotatebox{90}{\footnotesize $\Delta E_{LS}$ (eV)}
        \end{minipage}%
        \begin{minipage}[c]{\imgwd}
            \makebox[\imgwd][l]{%
                \includegraphics[width=\imgwd]{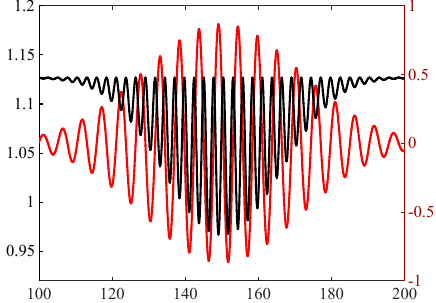}%
                \hspace{-\imgwd}%
                \raisebox{\dimexpr\imgh-0.85cm\relax}{\hspace{0.8cm}\textbf{(a)}}%
            }%
        \end{minipage}%
        \begin{minipage}[c]{\labelwd}
            \rotatebox{90}{\footnotesize $E$ (V/\AA)}
        \end{minipage}

        \par\vspace{2pt}
        \centering\footnotesize $t$ (fs)
        
        \vspace{0.0cm}

        \begin{minipage}[c]{\labelwd}
            \rotatebox{90}{\footnotesize $\mathcal{F}(\Delta E_{LS})$}
        \end{minipage}%
        \begin{minipage}[c]{\imgwd}
            \makebox[\imgwd][l]{%
                \includegraphics[width=\imgwd]{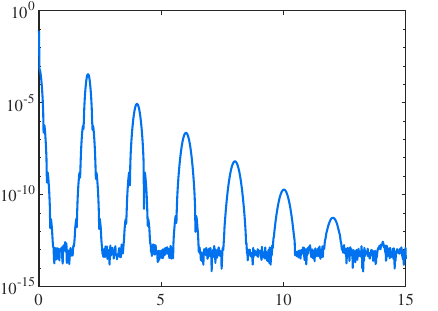}%
                \hspace{-\imgwd}%
                \raisebox{\dimexpr\imgh-0.85cm\relax}{\hspace{0.8cm}\textbf{(b)}}%
            }%
        \end{minipage}%
        \begin{minipage}[c]{\labelwd}
            \rotatebox{90}{  }

        \end{minipage}
        
        \par\vspace{2pt}
        \centering\footnotesize order
    \end{minipage}
    
    \caption{(Color online) (a) Time-dependent modulation of the energy gap $\Delta E_{{LS}}$ induced by the Lamb shift (black line), compared with the driving laser field $E(t)$ (red line). The modulation clearly oscillates at twice the frequency of the optical field. (b) Fourier spectrum $\mathcal{F}(\Delta E_{{LS}})$ of the Lamb-shift-induced energy gap modulation. The spectrum exhibits peaks exclusively at even multiples of the laser frequency $\omega$ (i.e., $0, \pm 2\omega, \pm 4\omega, \dots$), confirming the frequency-doubling nature of the Lamb shift. The results are obtained for $\omega_c = 10\Delta E$ and $\eta = 10^{-2}$.}
    \label{fig:lamb_shift_spectrum}
\end{figure}

This frequency-doubling characteristic is confirmed in Fig.~\ref{fig:lamb_shift_spectrum}, which shows the time-dependent modulation of the energy gap $\Delta E_{{LS}}$ induced by the Lamb shift and its Fourier spectrum, for $\eta = 10^{-2}$ and $\omega_c = 10\Delta E$ — the parameter regime where the order-dependent enhancement is most pronounced. The spectrum exhibits peaks exclusively at $0, \pm 2\omega, \pm 4\omega, \dots$, confirming that the Lamb shift modulates the system at even multiples of the driving frequency. Contributions from off-diagonal elements of the Lamb shift Hamiltonian are negligible.

To understand why higher harmonics are preferentially enhanced, we note that the Lamb shift acts as a periodic modulation of the energy levels at twice the laser frequency. Writing the Lamb shift Hamiltonian in the diagonal form \(H_{{LS}}(t) = -\sigma_z \sum_n b_n \cos(2n\omega t)\), where \(b_n\) are the Fourier coefficients of the Lamb-shift-induced energy modulation, we see that each frequency component \(2n\omega\) modulates the energy gap between the two levels. In the interaction picture defined by \(H_{S0} = -(\Delta E/2)\sigma_z\), this modulation imprints a periodic phase on the off-diagonal element of the density matrix, which can be expanded in a Fourier series with coefficients \(c_m\) (involving Bessel functions of the first kind). Consequently, the Fourier transform of the off-diagonal density matrix element satisfies
\begin{equation}
\mathcal{F}\bigl( [\rho_I(t)]_{12} \bigr)_n
= \sum_m c_m \, \mathcal{F}\bigl( [\rho'_I(t)]_{12} \bigr)_{n-2m}.
\label{eq:FT_relation}
\end{equation}
This relation has a transparent physical interpretation: the Lamb-shift modulation at frequency \(2m\omega\) shifts the \(n\)-th harmonic response to the \((n-2m)\)-th order, so that each harmonic order gains spectral weight through multiple sideband channels. The higher the harmonic order, the larger the number of such channels available, leading to a greater net enhancement. This sideband picture naturally explains why the Lamb-shift-induced gain grows with harmonic order, as observed in Fig.~\ref{fig:lamb}.

The comparison above explains the full results in Fig.~\ref{fig:full} as follows. For $\omega_c < \Delta E$, the suppression of harmonic intensity is dominated by the Lamb shift, with dissipation adding further reduction. For $\omega_c > \Delta E$, the situation splits by harmonic order. For low orders (5, 7, 9), the dissipative effect is relatively weak; the Lamb shift dominates, leading to a monotonic increase of the harmonic intensity with $\eta$. For high orders (11, 13, 15), dissipation becomes increasingly important as $\eta$ grows; its suppressing effect gradually overtakes the Lamb-shift-induced enhancement, resulting in the observed nonmonotonic behavior -- an initial rise followed by a decline.


We have studied how a dissipative environment modifies high-order harmonic generation in a driven two-level system, using a Lindblad master equation that separates the environmental influence into a Lamb shift and quantum dissipation. This decomposition, though standard in open quantum systems, turns out to be the key to understanding a non-trivial effect: the environment can either suppress or enhance harmonic emission, depending on whether the bath cutoff frequency falls below or above the level spacing. The enhancement is not a generic broadband gain; it is order-selective. Low harmonics respond monotonically to the coupling strength, while high harmonics display a nonmonotonic behavior—first rising, then falling—as dissipation begins to dominate over the Lamb-shift-induced modulation. What makes this selective response possible is the frequency-doubling nature of the Lamb shift. Because the instantaneous eigenfrequency of the driven two-level system depends on the square of the electric field, the Lamb shift oscillates at even multiples of the laser frequency. This creates sideband channels that preferentially couple to higher harmonics, explaining why they benefit more from the environmental coupling—up to the point where decoherence takes over.

Our results suggest a shift in perspective: a heat bath need not be viewed solely as a source of damping. Properly tuned, it can act as an active element that reshapes the harmonic spectrum in an order-dependent manner. The two control parameters in our model—the cutoff frequency and the coupling strength—correspond to experimentally accessible knobs, such as the choice of host lattice, nanostructuring, or dielectric environment. More broadly, this work provides a microscopic justification for moving beyond phenomenological dephasing times in HHG from complex system, and points toward a regime where environmental engineering becomes a viable strategy for controlling strong-field processes in quantum materials.

\section*{Acknowledgments}
This work was supported by the National Key R\&D Program of China (2022YFA1604301, 2024YFB3816300), National Natural Science Foundation of China (12425411, 12434013, 12450407, 12574371, 12304378, 12674425), the Natural Science Foundation of Jiangsu Province (BK20253027) and Fundamental Research Funds for the Central Universities (WZJC202602002).



\begin{thebibliography}{99}

\bibitem{McPherson1987}
A.~McPherson, G.~Gibson, H.~Jara, U.~Johann, T.~S.~Luk, I.~A.~McIntyre, K.~Boyer, and C.~K.~Rhodes,
J. Opt. Soc. Am. B \textbf{4}, 595 (1987).

\bibitem{Ferray1988}
M.~Ferray, A.~L'Huillier, X.~F.~Li, L.~A.~Lompre, G.~Mainfray, and C.~Manus,
J. Phys. B: At. Mol. Opt. Phys. \textbf{21}, L31 (1988).

\bibitem{Corkum1993}
P.~B.~Corkum,
Phys. Rev. Lett. \textbf{71}, 1994 (1993).

\bibitem{Lewenstein1994}
M.~Lewenstein, Ph.~Balcou, M.~Yu.~Ivanov, A.~L'Huillier, and P.~B.~Corkum,
Phys. Rev. A \textbf{49}, 2117 (1994).

\bibitem{Ghimire2011}
S.~Ghimire, A.~D.~DiChiara, E.~Sistrunk, P.~Agostini, L.~F.~DiMauro, and D.~A.~Reis,
Nat. Phys. \textbf{7}, 138 (2011).

\bibitem{Wiechmann2025}
F.-E.~Wiechmann, S.~Sch\"opa, L.~Bielke, S.~Rindelhardt, S.~Patchkovskii, F.~Morales, M.~Richter, D.~Bauer, and F.~Fennel,
Nat. Commun. \textbf{16}, 9890 (2025).

\bibitem{Spintextures2025}
F.~Gabriele, C.~Ortix, M.~Cuoco, and F.~Forte,
Phys. Rev. B \textbf{112}, 155421 (2025).

\bibitem{Monolayer2025}
A.~M.~Koushki,
Appl. Phys. B \textbf{131}, 74 (2025).

\bibitem{Luu2018}
T.~T.~Luu, Z.~Yin, A.~Jain, T.~Gaumnitz, Y.~Pertot, J.~Ma, and H.~J.~W\"orner,
Nat. Commun. \textbf{9}, 3723 (2018).

\bibitem{Mondal2025}
A.~Mondal, O.~Neufeld, T.~Bal{\v c}iūnas, B.~Waser, S.~M{\"u}ller, M.~Rossi, Z.~Yin, and A.~Rubio,
Nat. Photon. \textbf{20}, 216 (2026).

\bibitem{Kim2022}
V.~V.~Kim, R.~A.~Ganeev, S.~R.~Konda, G.~S.~Boltaev, I.~B.~Sapaev, and A.~S.~Alnaser,
Sci. Rep. \textbf{12}, 9128 (2022).

\bibitem{Ganeev2023}
R.~A.~Ganeev,
Appl. Phys. B \textbf{129}, 17 (2023).

\bibitem{Vismarra2023}
F.~Vismarra, D.~Mocci, L.~Colaizzi, M.~F.~Gal{\'a}n, V.~W.~Segundo, R.~Boyero-Garc{\'i}a, J.~Serrano,
E.~Conejero~Jarque, M.~Pini, L.~Mai, Y.~Wu, M.~Reduzzi, M.~Lucchini, H.~J.~W\"orner,
C.~L.~Arnold, J.~San~Rom{\'a}n, C.~Hern{\'a}ndez-Garc{\'i}a, M.~Nisoli, and R.~Borrego-Varillas,
in \emph{2023 Conference on Lasers and Electro-Optics Europe \& European Quantum Electronics Conference (CLEO/EQEC)},
IEEE, Munich, Germany, paper cf\_1\_3 (2023).

\bibitem{Guo2026}
Z.~Guo, X.~Tang, C.~Zhang, B.~Lu, S.~Wu, T.~Wen, Z.~Chen, B.~Wang, X.~Li, C.~D.~Lin, and C.~Jin,
Photon. Res. \textbf{14}, 1832 (2026).

\bibitem{Parks2025}
A.~M.~Parks and M.~Kolesik,
Opt. Express \textbf{33}, 13986 (2025).

\bibitem{Banks2017}
H.~B.~Banks, Q.~Wu, D.~C.~Valovcin, S.~Mack, A.~C.~Gossard, L.~Pfeiffer, R.-B.~Liu, and M.~S.~Sherwin,
Phys. Rev. X \textbf{7}, 041042 (2017).

\bibitem{Rutledge2021}
J.~Rutledge, A.~Catanese, D.~D.~Hickstein, S.~A.~Diddams, T.~K.~Allison, and A.~S.~Kowligy,
J. Opt. Soc. Am. B \textbf{38}, 2252 (2021).

\bibitem{Di2023}
Y.~Di, Z.~Zuo, D.~Peng, D.~Luo, C.~Gu, and W.~Li,
Photonics Res. \textbf{11}, 1373 (2023).

\bibitem{Gorlach2020}
A.~Gorlach, O.~Neufeld, N.~Rivera, O.~Cohen, and I.~Kaminer,
Nat. Commun. \textbf{11}, 4598 (2020).

\bibitem{Stammer2024}
P.~Stammer, J.~Rivera-Dean, A.~S.~Maxwell, T.~Lamprou, J.~Arg\"uello-Luengo, P.~Tzallas, M.~F.~Ciappina, and M.~Lewenstein,
Phys. Rev. Lett. \textbf{132}, 143603 (2024).

\bibitem{jiangCPL}
S.~Jiang and K.~Dorfman,
Chin. Phys. Lett. \textbf{43}, 050401 (2026).

\bibitem{TopoMeta2025}
M.~S.~Vitiello et al.,
Light Sci. Appl. \textbf{14}, 337 (2025).

\bibitem{Brown2024}
G.~G.~Brown, \'A.~Jim\'enez-Gal\'an, R.~E.~F.~Silva, and M.~Ivanov,
Phys. Rev. Research \textbf{6}, 043005 (2024).

\bibitem{Bae2026}
G.~Bae, Y.~Kim, and J.~D.~Lee,
Advanced Science \textbf{13}, 2522729 (2026).

\bibitem{Hu2024}
S.-Q.~Hu, H.~Zhao, X.-B.~Liu, Q.~Chen, D.-Q.~Chen, X.-Y.~Zhang, and S.~Meng,
Phys. Rev. Lett. \textbf{133}, 156901 (2024).

\bibitem{Herling2026}
A.~Herling and O.~Neufeld,
J. Phys. Chem. Lett. \textbf{17}, 7936 (2026).

\bibitem{Liu2025}
H.~Liu, Z.~Wang, and R.~Lu,
Sci. China-Phys. Mech. Astron. \textbf{69}, 214231 (2026).

\bibitem{Frontiers2025}
M.~F.~Ciappina,
Adv. Phys. \textbf{74}, 1 (2025).

\bibitem{Discreteness2023}
M.~Kolesik and J.~V.~Moloney,
Phys. Rev. B \textbf{108}, 115433 (2023).

\bibitem{UltrafastDephasing2023}
G.~G.~Brown, \'A.~Jim\'enez-Gal\'an, R.~E.~F.~Silva, and M.~Ivanov,
J. Opt. Soc. Am. B \textbf{41}, B40 (2024).

\bibitem{Boroumand2025}
N.~Boroumand, A.~Parks, Y.~Liu, T.~Brabec, and L.~Wang,
Rep. Prog. Phys. \textbf{88}, 070501 (2025).

\bibitem{Reentrance2026}
W.~Hogger, A.~Riedel, D.~Roy, A.~Knothe, C.~Gorini, J.-D.~Urbina, and K.~Richter,
Phys. Rev. B \textbf{113}, 045418 (2026).

\bibitem{Decoherence2021}
G.~Wang and T.-Y.~Du,
Phys. Rev. A \textbf{103}, 063109 (2021).

\bibitem{Incoherent2023}
K.~Nagai, K.~Uchida, S.~Kusaba, T.~Endo, Y.~Miyata, and K.~Tanaka,
Phys. Rev. Research \textbf{5}, 043127 (2023).

\bibitem{Kruchinin2019}
S.~Y.~Kruchinin,
Phys. Rev. A \textbf{100}, 043839 (2019).

\bibitem{Du2022}
T.-Y.~Du and C.~Ma,
Phys. Rev. A \textbf{105}, 053125 (2022).

\bibitem{Trajectory2024}
Y.~Wang, T.~Shao, X.~Li, Y.~Liu, P.~Jiang, W.~Zheng, L.~Zhang, X.-B.~Bian, Y.~Liu, Q.~Gong, and C.~Wu,
Opt. Express \textbf{31}, 3379 (2023).

\bibitem{SpinHHG2024}
N.~Moharrami~Allafi, M.~H.~Kolodrubetz, M.~Bukov, V.~Oganesyan, and M.~Yarmohammadi,
Phys. Rev. Lett. \textbf{133}, 167401 (2024).

\bibitem{Lindblad1976}
G.~Lindblad,
Commun. Math. Phys. \textbf{48}, 119 (1976).

\bibitem{Gorini1976}
V.~Gorini, A.~Kossakowski, and E.~C.~G.~Sudarshan,
J. Math. Phys. \textbf{17}, 821 (1976).

\bibitem{Breuer2002}
H.-P.~Breuer and F.~Petruccione,
(Oxford University Press, Oxford, 2002).

\bibitem{Leggett1987}
A.~J.~Leggett, S.~Chakravarty, A.~T.~Dorsey, M.~P.~A.~Fisher, A.~Garg, and W.~Zwerger,
Rev. Mod. Phys. \textbf{59}, 1 (1987).

\bibitem{Caldeira1983}
A.~O.~Caldeira and A.~J.~Leggett,
Physica A \textbf{121}, 587 (1983).

\bibitem{Gardiner2000}
C.~W.~Gardiner and P.~Zoller,
2nd ed. (Springer, Berlin, 2000).

\bibitem{Supplemental}
See Supplemental Material at [URL inserted by publisher] for the detailed derivation of the Lindblad master equation and the explicit forms of the bath correlation functions.

\end{thebibliography}
\end{document}